\documentclass[prl,twocolumn,amssymb,showpacs,amsmath,nobibnotes,superscriptaddress,floatfix, aps]{revtex4}
\usepackage{graphicx}
\usepackage{bm}

\newcommand{\be}{\begin{equation}}
\newcommand{\ee}{\end{equation}}
\newcommand{\bea}{\begin{eqnarray}}
\newcommand{\eea}{\end{eqnarray}}

\newcommand{\order}{{\cal O}}
\newcommand{\fig}[1]{Fig.~\ref{#1}}
\newcommand{\tble}[1]{Table~\ref{#1}}
\newcommand{\alphamsb}{\alpha_{\overline{\mathrm{MS}}}}
\newcommand{\alphav}{\alpha_V}
\newcommand{\eq}[1]{Eq.~(\ref{#1})}
\begin{document}
\title{Accurate Determinations of $\alpha_s$ from Realistic Lattice QCD}
\author{Q.~Mason}
\affiliation{Department of Applied Mathematics and Theoretical Physics,
University of Cambridge, Cambridge, United Kingdom}
\author{H.~D.~Trottier}
\affiliation{Physics Department, Simon Fraser University, Vancouver, British Columbia, Canada}
\author{C.~T.~H.~Davies}
\affiliation{Department of Physics and Astronomy, University of Glasgow, Glasgow, United Kingdom}
\author{K.~Foley}
\affiliation{Laboratory for Elementary-Particle Physics, Cornell University, Ithaca, NY 14853}
\author{A.~Gray}
\affiliation{Physics Department, The Ohio State University, Columbus, OH 43210}
\author{G.~P.~Lepage}
\author{M.~Nobes}
\affiliation{Laboratory for Elementary-Particle Physics, Cornell University, Ithaca, NY 14853}
\author{J.~Shigemitsu}
\affiliation{Physics Department, The Ohio State University, Columbus, OH 43210}
\collaboration{HPQCD and UKQCD Collaborations}
\noaffiliation

\date{4 June 2005}
\pacs{11.15.Ha,12.38.Aw,12.38.Gc}
\begin{abstract}
We obtain a new value for the QCD coupling constant by combining lattice QCD simulations with experimental data for hadron masses. Our lattice analysis is the first to: 1) include vacuum polarization effects from all three light-quark flavors (using MILC configurations); 2)  include third-order terms in perturbation theory; 3)  systematically estimate fourth and higher-order terms; 4) use an unambiguous  lattice spacing; and 5) use an $\order(a^2)$-accurate QCD action. We use 28~different (but related) short-distance quantities to obtain
$\alpha_{\overline{\mathrm{MS}}}^{(5)}(M_Z) = 0.1170(12)$.
\end{abstract}
\maketitle

An accurate value for the coupling constant~$\alpha_s$ in quantum chromodynamics (QCD) is important both for high-energy phenomenology, and as an input for possible theories beyond the Standard Model. Numerical simulations of QCD using lattice techniques, when combined with experimental data for hadron masses, have provided some of the most accurate values for the coupling constant\,\cite{alpha-papers}. The precision of these determinations has been limited, however, by two factors. One was our inability to include the effects of realistic light-quark vacuum polarization in QCD simulations. The other limitation was the lack of third and higher order terms in the perturbative expansions used to extract~$\alpha_s$. In this paper we present the first lattice QCD determination of the coupling constant that includes realistic vacuum polarization effects from all three light quarks, and perturbation theory through third order, with systematic estimates of fourth order and beyond. Consequently, our final results are, by far, the most accurate from lattice QCD and among the most accurate from any method. This work uses gluon configurations from the MILC collaboration\,\cite{MILCpaper}, and builds on a joint effort by several groups\,\cite{ratio-paper}.

Effects from light-quark vacuum polarization are quantitatively important, but also very costly to simulate. Previous simulations included contributions from only $u$~and $d$~quarks, no $s$~quarks, and used quark masses that were 10~times too large or larger. Our analysis includes effects from all three light quarks, with much smaller $u$~and $d$~masses\,---\,so small that our results become effectively mass-independent. This is possible because of a new discretization of the light-quark action\,\cite{ratio-paper}. Heavy-quark polarization is negligible and is ignored here\,\cite{nobes-paper}. 

The Lorentz-noninvariant ultraviolet regulator greatly complicates high-order perturbation theory in lattice QCD. To manage this complexity, we automated the generation of Feynman integrands using computers, and evaluated the Feynman integrals numerically on large-scale parallel computers. These techniques allowed us to evaluate perturbative coefficients through third order\,\cite{loop-paper}.

To extract the coupling constant from our lattice QCD simulation, we (with our collaborators) first tuned the theory's five parameters to reproduce experiment for five well-measured quantities; the details are in~\cite{ratio-paper}. We used lattices that were approximately 2.5\,fm on a side with three different lattice spacings~$a$, where $a^{-1}$ was 1.144(31), 1.596(30), and 2.258(32)\,GeV. The $s$~quark masses we used for each of the three lattice spacings were $0.082/a$, $0.05/a$, and $0.031/a$, respectively. We used $u$ and $d$ masses as small as $m_s/5$, except on the coarsest lattice where we used~$m_s/10$. The gluon configurations were produced using an improved gluon action and the new light-quark action. Our coupling-constant analysis is the first to use $\order(a^2)$-accurate actions.

The lattice spacing~$a$ is one of the five simulation parameters, and the most important in our analysis because it sets the simulation's mass scale. In our earlier $\alpha_s$~analyses,   we set the lattice spacing by comparing a simulated $\Upsilon$ mass splitting (\emph{e.g.}, $\Upsilon^\prime-\Upsilon$) with experiment. Here we continue this practice, but, for the first time, the lattice spacings derived from our $\Upsilon$~splitting have been shown to agree with spacings derived from a wide variety of other physical quantities: ten in all, including the pion and kaon leptonic decay constants, the $B_s$~mass, and the $\Omega$~baryon mass\,\cite{ratio-paper,doug-ratio-paper}. All of these different quantities agree on the lattice spacing to within~1.5--3\%.

Having an accurately tuned simulation of QCD, we used it to compute nonperturbative values for a variety of short-distance quantities, each of which has a perturbative expansion of the form
\be
\label{pert-exp}
Y = \sum_{n=1}^{\infty} c_n\,\alphav^n(d/a)
\ee
where $c_n$ and $d$ are dimensionless $a$-independent constants,  and $\alphav(d/a)$ is the (running) QCD coupling constant, for $n_f=3$ flavors, in the $V$~scheme\,\cite{tadpole-improvement-paper,alphaV-paper}. Given the coefficients~$c_n$, we determine $\alphav(d/a)$ such that the perturbative formula for $Y$ reproduces the nonperturbative value from the simulation. 

We computed $c_n$ for $n\le3$ using Feynman diagrams. We estimated higher-order coefficients by simultaneously fitting results from different lattice spacings to the same perturbative formula. This is possible because the coupling $\alphav(d/a)$ changes value with different lattice spacings~$a$:
\be \label{evol-eq}
q^2\,\frac{d\alphav(q)}{dq^2} = - \beta_0 \alphav^2 - \beta_1 \alphav^3
-\beta_2 \alphav^4 -\beta_3\alphav^5 - \cdots
\ee
where the $\beta_i$ are constants\,\cite{alphaV-paper}. We parameterize the running coupling in our fits by its value at a specific scale\,---\,$\alpha_0\equiv\alphav(7.5\,\mathrm{GeV})$\,---\,and integrate \eq{evol-eq} numerically to obtain values at other scales.

We used a constrained fitting procedure, based upon Bayesian methods, for our fits\,\cite{bayes-paper}. Given simulation results $Y_i\pm\sigma_{Y_i}$ for three different lattice spacings~$\overline{a_i}\pm\sigma_{a_i}$, we minimized an augmented $\chi^2$ function,
\bea
\chi^2 &\equiv& \sum_{i=1}^{3} 
\frac{\left(Y_i - \sum_n c_n\,\alphav^n(d/a_i)\right)^2}{\sigma_{Y_i}^2}
+ \sum_{n=1}^{10} \frac{(c_n-\overline{c_n})^2}{\sigma_{c_n}^2}
\nonumber \\
&&+ \frac{\left(\log(\alpha_0)-\overline{\log(\alpha_0)}\right)^2}{\sigma_{\log(\alpha_0)}^2}
+ \sum_{i=1}^3 \frac{(a_i -\overline{a_i})^2}{\sigma_{a_i}^2},
\eea
by varying  $\alpha_0$, and the $c_n$ and $a_i$. Terms after the first in $\chi^2$ are ``priors'' that constrain the fit parameters to a reasonable range. The fits explored values of $c_n$, for example, centered around $\overline{c_n}$ with a range specified by $\sigma_{c_n}$. For $n\le3$, we set $\overline{c_n}$ to the values obtained from our numerical evaluations of the relevant Feynman diagrams, with $\sigma_{c_n}$ equal to the uncertainties in those evaluations. We set $\overline{c_n}=0$ for $4\le n\le10$, and $\sigma_{c_n}=\sigma_c$ where $\sigma_c$ was determined using the Empirical Bayes procedure described in~\cite{bayes-paper}. (Typically this procedure set $\sigma_c$ somewhat larger than the optimal value found for $|c_4|$.) We ignored terms with~$n>10$. The prior constraint on the coupling constant, $\alpha_0$, was $0.20^{+0.20}_{-0.10}$, or equivalently $-1.6\pm0.7$ for $\log(\alpha_0)$; it had negligible impact on the fits. 

The simplest short-distance quantities to simulate are vacuum expectation values of Wilson loop operators:
\be
W_{mn} \equiv \mbox{$\frac{1}{3}$}\,\langle0|\, \mathrm{Re\,Tr}\,\mathrm{P}\,\mathrm{
e}^{-ig\oint_{nm}
\!A\cdot dx}\, |0\rangle,
\ee
where ${\rm P}$~denotes path ordering, $A_\mu$~is the QCD vector
potential, and the integral is over a closed
${ma\!\times\!na}$~rectangular path. Wilson loops are perturbative when~$ma$ and~$na$ are small. We computed perturbative coefficients  for six small loops\,\cite{loop-paper}, ``measured'' them nonperturbatively in simulations with each of our three lattice spacings, and did fits to perturbation theory for each loop. We also evaluated Wilson loops for two non-planar paths\,\cite{loop-paper}:
\vspace{-3ex}
\be
\mathrm{BR} = 
\begin{picture}(60,30)(0,15)
  \put(10,10){\vector(0,1){12.5}}
  \put(10,10){\line(0,1){20}}
  \put(10,30){\vector(2,1){10}}
  \put(10,30){\line(2,1){15}}
  \put(25.2,37.6){\vector(1,0){12.5}}
  \put(25.2,37.6){\line(1,0){20}}
  \put(45.2,37.6){\vector(0,-1){12.5}}
  \put(45.2,37.6){\line(0,-1){20}}
  \put(45.2,17.6){\vector(-1,0){12.5}}
  \put(45.2,17.6){\line(-1,0){20}}
  \put(25.2,17.6){\vector(-2,-1){10}}
  \put(25.2,17.6){\line(-2,-1){15}}
\end{picture}
\quad
\mathrm{CC} =
\begin{picture}(60,30)(0,15) 
  \put(10,10){\vector(0,1){12.5}}
  \put(10,10){\line(0,1){20}}
  \put(10,30){\vector(2,1){10}}
  \put(10,30){\line(2,1){15}}
  \put(25.2,37.6){\vector(1,0){12.5}}
  \put(25.2,37.6){\line(1,0){20}}
  \put(45.2,37.6){\vector(0,-1){12.5}}
  \put(45.2,37.6){\line(0,-1){20}}
  \put(45.2,17.6){\vector(-2,-1){10}}
  \put(45.2,17.6){\line(-2,-1){15}}
  \put(30,10){\vector(-1,0){12.5}}
  \put(30,10){\line(-1,0){20}}
\end{picture}.
\ee

The fits revealed that high-order coefficients in the Wilson loop expansions are larger than we expected: for example, we find
\bea
\log W_{11} &=& -3.068\,\alphav(3.33/a)\, 
\left(1 - 1.068\,\alphav   \right.   
 \\ \nonumber
&&\left. +1.69(4)\,\alphav^2  - 5(2)\,\alphav^3 - 1(6)\,\alphav^4\,\cdots\right); 
\\
\log W_{12} &=& -5.551\,\alphav(3.00/a)\, 
\left(1 - 0.858\,\alphav   \right.   
 \\ \nonumber
&&\left. +1.72(4)\,\alphav^2  - 5(2)\,\alphav^3 - 1(6)\,\alphav^4\,\cdots\right).
\eea
The large $5\alphav^3$ corrections are needed if perturbation theory is to agree with simulation results for all three lattice spacings. The coupling $\alphav(3.33/a)$ ranges between 0.21 and 0.29 for our lattice spacings, so $5\alphav^3$ is 5--12\% of the full result. Each Wilson loop we examined had corrections of order this size.

\begin{figure}
\begin{center}
\includegraphics[scale=1.0]{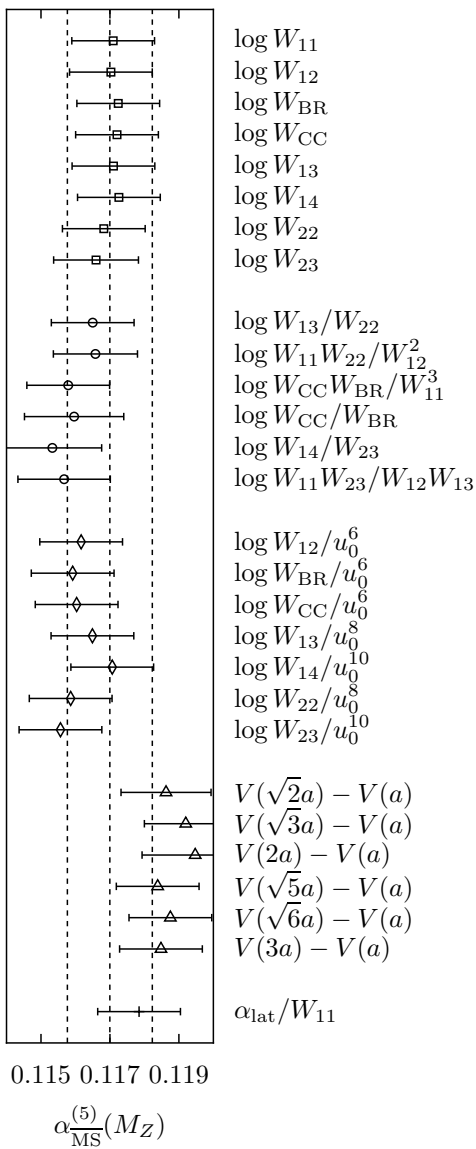} 
\end{center}
\caption{Values for the 5-flavor $\alphamsb$ at the $Z$~mass from each short-distance quantity. The dashed lines indicate our final result, 0.1170(12) ($\chi^2$ per data point is~0.77).}
\label{answer-plot}
\end{figure}

These large coefficients reduce the accuracy of our final results. There are two ways to reduce the size of these coefficients. One is to ``tadpole improve'' $W_{mn}$ by dividing by $u_0^{2(n+m)}$ where $u_0\equiv(W_{11})^{1/4}$\,\cite{tadpole-improvement-paper}. The other is to examine Creutz ratios of the loops rather than the loops themselves\,\cite{tadpole-improvement-paper}. Each procedure significantly reduces the high-order coefficients we obtain when we refit to results from our three lattice spacings: for example,
\bea
\log\left(\frac{W_{12}}{u_0^6}\right) &=& 0.949\,\alphav(1.82/a)\,
\left(1 + 0.160(2)\,\alphav \right.
\\ \nonumber
&& \left. - 0.54(8)\,\alphav^2 -2(1)\,\alphav^3 - 0(2)\,\alphav^4\,\cdots\right);
\\
\log\left(\frac{W_{13}}{W_{22}}\right) &=& -1.323\,\alphav(1.21/a)\,
\left(1 - 0.39(1)\,\alphav \right. \\ \nonumber
&&\left. + 0.3(2)\,\alphav^2 - 2(1)\,\alphav^3 + 0(2)\,\alphav^4\,\cdots\right).
\eea
These expansions are typical of the 7 tadpole-improved loops and 6 Creutz ratios that we examined. Each has smaller $\alphav^3$ coefficients, but also significantly smaller scales for the $\alphav$s. Over our range of lattice spacings, $\alphav(1.21/a)$, for example, ranges between~0.33 and~0.68, and therefore $2\alphav^3$ is 7--60\% of the final result depending upon the lattice spacing. Consequently results from these quantities are not significantly more accurate than those from Wilson loops. Results from the coarsest lattices, with large~$\alphav$s, carry the least weight in our fits.

We also examined the static-quark potential, which is perturbative at short distances. The continuum potential has a particularly simple form in the $V$~scheme:
\be
V(r) = -C_F\,\frac{\alphav(0.5614/r)}{r}\,\left(1 + \frac{\beta_0^2}{48}\,\alphav^2+\cdots\right)
\ee
where $C_F=4/3$ and $\beta_0=11-2n_f/3$. On the lattice we examined the quantity $V(r)-V(a)$ since lattice artifacts cancel almost completely in the difference. We computed and removed the small residual lattice artifacts in $V(r)-V(a)$ through second order in perturbation theory, and fit the resulting potential with the continuum formula; higher-order lattice artifacts are negligible here. Continuum perturbation theory for $V(r)-V(a)$ becomes nonanalytic, however, in fourth order, with the appearance of terms proportional to $\alphav^4\log(\alphav)$\,\cite{dine-paper}. $\log(\alphav)$ is small for our range of $\alphav$s, so we see no evidence of it in our fits. Nevertheless, the presence of such terms suggests that our results from the potential may not be as reliable as those from our other quantities. We limited our analysis to $r\le3a$ as otherwise the $\alphav$~scales become too small ($<1/a$). For the same reason, we discarded results for the potential from the coarsest lattice.

\begin{figure}
\begin{center}
\includegraphics[scale=1.0]{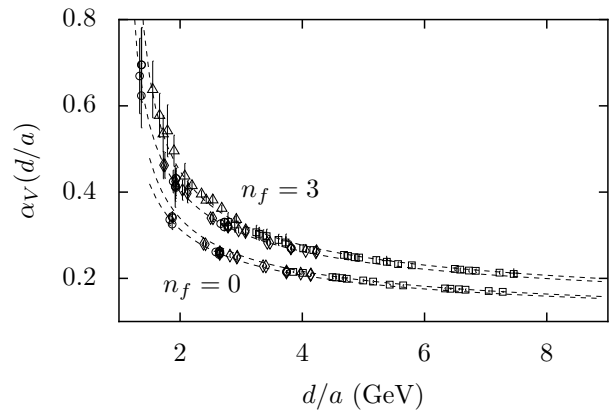} 
\end{center}
\caption{Values for $\alphav$ versus $d/a$ (\eq{pert-exp}) from each short-distance quantity at each lattice spacing, with (top) and without (bottom) light-quark vacuum polarization. The dashed lines show predictions from \eq{evol-eq}  assuming $\alpha_V(7.5\,\mathrm{GeV})$ is $0.2082(40)$ and $0.1645(14)$ for $n_f=3$ and~0, respectively.}
\label{alpha-plot}
\end{figure}

Finally, we extracted the coupling directly from the tadpole-improved bare lattice coupling, $\alpha_\mathrm{lat}/W_{11}$, which, like Wilson loops, has large fourth-order coefficients.

We extracted values for the 3-flavor coupling, $\alpha_0\equiv\alpha_V(7.5\,\mathrm{GeV})$, 
from fits to each of our 28 short-distance quantities. To facilitate comparison with other determinations, we converted our results
from the $V$~scheme to the $\overline{\mathrm{MS}}$~scheme\,\cite{alphaV-paper}, added $c$~and $b$~quark vacuum polarization (perturbatively\,\cite{threshold-paper}, using quark masses of 1.25(10) and 4.25(15)\,GeV\,\cite{PDG-paper}), and evolved to the $Z$~mass. The results from the different quantities are shown in~\fig{answer-plot}. 

While they are derived from the Wilson loops, our Creutz ratios and tadpole-improved loops provide coupling-constant information that is largely independent of that coming from the loops. This is because the highly ultraviolet contributions that dominate the loops largely cancel in the other quantities, making the latter far more infrared (\emph{c.f.}, $(d/a)$s for loops and ratios). Our 28 separate determinations of the scale parameter 
probe a wide range of different length scales, have very different sensitivities to potential nonperturbative errors, and, as we have discussed, have very different perturbative expansions. The agreement, to within our errors, of all 28 determinations is strong evidence that we have correctly identified and controlled the various systematic errors that could have affected our analysis. 
 
The weighted average of our 28~determinations gives a composite result of
\be
\alpha_{\overline{\mathrm{MS}}}^{(5)}(M_Z) = 0.1170(12),
\ee
or, equivalently,
\be
\alpha_V^{(3)}(7.5\,\mathrm{GeV}) = 0.2082(40).
\ee
Our error estimate here is that of a typical entry in the plot; combining our results does not reduce errors because most of the uncertainty in each result is systematic.

Our composite value for the coupling constant agrees well with the current Particle Data Group world average value of 0.1187(20)\,\cite{PDG-paper}, but is somewhat more accurate. It also agrees within errors with our previous results\,\cite{alpha-papers}, and with the preliminary result of the present analysis presented in~\cite{ratio-paper}, where we quoted larger uncertainties because we lacked values for the~$c_3$s.

Realistic vacuum polarization is critical to our result. Redoing our simulations and analysis, but with no light-quark vacuum polarization, gives a coupling of~0.0900(4) rather than~0.1170(12). Evolving to $M_Z$ increases the difference between 0~and 3~light-quark flavors, but the couplings are still $10\sigma$~apart at 7.5~GeV. This is evident in \fig{alpha-plot} where we plot the $\alphav(d/a)$s extracted from our fits, with and without vacuum polarization, for each quantity and for each of our lattice spacings separately, but using the $c_n$s from our simultaneous fits to all lattice spacings.

\begin{table}
\begin{center}
\begin{tabular}{rccc}
\\
	&	$\log W_{11}$	& $\log W_{13}/W_{22}$ &  $V(\sqrt{2}a)-V(a)$ \\ \hline
$a^{-1}$					        & 0.0007 & 0.0010 &  0.0010  \\
$c_1\ldots c_3$ 			        & 0.0001 & 0.0004 &  0.0004  \\
$c_n$ for $n\ge4$			        & 0.0008 & 0.0005 &  0.0004  \\
$V\to\overline{\mathrm{MS}}\to M_Z$ & 0.0001 & 0.0001 &  0.0001  \\
condensate					        & 0.0002 & 0.0001 &  0.0001  \\
$m_u$, $m_d$, $m_s$			        & 0.0004 & 0.0001 &  0.0001  \\
$m_c$, $m_b$				        & 0.0002 & 0.0002 &  0.0002  \\
simulation errors			        & 0.0000 & 0.0000 &  0.0002  \\
\hline                        
total	uncertainty			        & 0.0012 & 0.0012 &  0.0012  \\
\end{tabular}
\end{center}
\caption{Sources of the uncertainties in our final determinations of the coupling $\alpha_{\overline{\mathrm{MS}}}^{(5)}(M_Z) =0.1170(12)$.}
\label{error-table}
\end{table}

The various sources of uncertainty for different quantities are elaborated in \tble{error-table}. The dominant errors come from three sources. First is the uncertainty in the inverse lattice spacing~$a^{-1}$, which includes both statistical and estimated finite-$a$ errors in the simulated upsilon splitting (0.5--2\% depending upon~$a$\,\cite{ups-scaling-paper}). Second are residual uncertainties in the parameters $c_1\ldots c_3$ from the numerical calculation of these coefficients. The final large source of uncertainty is due to uncertainties in the coefficients beyond third order. This error is greatly reduced because we fit simultaneously to three lattice spacings; fitting with just a single lattice spacing, as is usually done, would give errors 2--5~times larger. We allowed for possible effects from nonperturbative gluon and quark condensates\,\cite{condensate-paper}, but these are negligible.
Uncertainties in the $c$ and $b$ masses, and Monte Carlo simulation errors in the loop values are negligible. We corrected for the errors in $u$, $d$ and $s$ sea-quark masses by redoing our entire analysis (loops and lattice spacings) with larger masses and extrapolating linearly. This is also negligible; we include an uncertainty in~\tble{error-table} equal to the correction.

Our coupling-constant analysis uses the most realistic QCD simulation to date, with, for the first time, vacuum polarization contributions from all three light-quarks, quark and gluon actions corrected through~$\order(a^2)$, and extensive evidence that both the heavy- and light-quark sectors of the theory have been accurately simulated\,\cite{ratio-paper}. It is the first to use not only third-order accurate perturbation theory, but also systematic estimates of fourth order and higher. Our final results come from 28 different short-distance quantities, covering almost an order of magnitude in energy scales. The agreement between our results and the current world average demonstrates that the QCD of confinement is the same theory as the QCD of jets; lattice QCD is full QCD, encompassing both its perturbative and nonperturbative aspects. 

This work was supported by NSERC (Canada), PPARC (UK), the NSF and the DOE (USA), and by SciDAC computing allocations at FNAL. We also thank Steve Gottlieb, Bob Sugar, Doug Toussaint, the MILC collaboration, Matthew Wingate, and Peter Boyle.

\end{document}